\documentclass[aps,prb,singlecolumn]{revtex4}
\usepackage{graphicx}
\begin{document}
\title{Wannier Orbital Overlap Population (WOOP), Wannier Orbital Position Population (WOPP)
and the Origin of Anomalous Dynamical Charges}
\author{Joydeep Bhattacharjee and Umesh V Waghmare\\
\small{\textit{Theoretical Sciences Unit}}\\
\small{\textit{Jawaharlal Nehru Centre for Advanced Scientific Research}}\\
\small{\textit{Jakkur PO, Bangalore 560 064, India}}}
\date{November 1, 2007}
\begin{abstract}
Most $d^0$ transition metal (TM) oxides exhibit anomalously large Born dynamical charges
associated with off-centering or motion of atoms along the TM-O chains. To understand
their chemical origin, we introduce ``Wannier Orbital Overlap Population" (WOOP) and
``Wannier ond Orbital Position Population" (WOPP) based on the Wannier function based description of 
electronic structure obtained within first-principles density functional theory.
We apply these concepts in a precise analysis of anomalous dynamical charges in 
PbTiO$_3$, BaTiO$_3$ and BaZrO$_3$ in the cubic perovskite structure. Determining contributions 
of different atomic orbitals to the dynamical charge and their break-up into local polarizability, 
charge transfer and covalency, we find that $p$ orbitals of oxygen perpendicular to the 
-TM-O- chain contribute most prominently to the anomalous charge, by facilitating a 
transfer of \textit{tiny} electronic charge through one unit cell from one TM atom to the next.
Our results explain why the corner-shared linkage of TMO$_6$ octahedra, as in 
the perovskite structure, is ideal for large dynamical charges and hence for 
ferroelectricity.
\end{abstract}
\maketitle
\section{Introduction}

Transition metal oxides (TMOs) exhibit a rich variety of structural, electric and magnetic 
properties\cite{TMO1,TMO2}, owing mostly to the $d$ electronic states of the transition metal.
Partially occupied $d$ states of a TM in its oxide result in different possibilities of 
magnetic ordering and properties, where as the TMOs with unoccupied $d$ states usually
are non-magnetic insulators with structural and dielectric properties of technological 
importance\cite{Diel}. Most ferroelectric (ABO$_3$) oxides in the perovskite 
structure\cite{devices} have a $d^0$ transition metal at the $B$-site, and interaction between
$d$ states of TM with $p$ states of O are known to be crucial to their properties\cite{Cohen}.
Many binary $d^0$ TM oxides such as TiO$_2$ and ZrO$_2$ hold promise as high-$k$ dielectric 
materials\cite{wood} essential to the future ultra-thin semiconductor devices. Understanding how precisely
the $d^0$ state of a TM atom and its interaction with oxygen gives rise to these properties is 
fundamentally interesting and will also help in design of better materials.

Dielectric properties are intimately related to electric polarization (dipole moment 
per unit volume) of insulators. It is spontaneously present 
in a ferroelectric material\cite{devices} even in the absence of external fields, 
and its coupling with structure, electric and elastic fields are central to technologically important 
properties of these materials. Dynamical charges, known as Born effective charges(BEC), 
yield polarization induced due to small displacements of atoms from their equilibrium positions.
Anomalous charge is evaluated as the difference between the dynamical charge associated with an atom 
and the nominal charge represented by it's oxidation state.
They are also the coupling constants between atomic displacements and electric
field, and are relevant to the properties of materials in two important ways: (a) they often
are anomalously large and indicators\cite{ZKV, UVW03} of the ferroelectric structural instabilities 
or low frequency (soft) polar phonon modes and (b) static dielectric (piezoelectric) response 
depends quadratically (linearly) on the effective charges and is dominated by the 
softest phonons indicated by the anomalous effective charges\cite{Diel}.

First-principles calculations based on density functional theory (DFT) have been
extensively used in determining dielectric\cite{Diel} and ferroelectric\cite{VanOp}
properties of the $d^0$ transition metal oxides. Within first-principles DFT framework, 
BECs are commonly calculated using two methods: (a) Berry phase 
based expression of polarization and (b) DFT-linear response. Both approaches are 
based in the $k-$space and do not readily yield a picture in real-space of the flow of 
electronic charge in terms of bonding. Band-by-band decomposition\cite{GhosezGonze} of 
effective charges using the linear response method was used to estimate contributions of 
different atomic orbitals to their anomalously large values, highlighting the role
of hybridization between $d$ states of TM with $p$ states of oxygen. 
The band-by-band decomposition\cite{Dvw} of BECs associated with the Ba, Ti and O atoms in BaTiO$_3$,
in groups of bands corresponding to different maximally localized Wannier functions\cite{MV}, was
found to be in good agreement with that obtained using the linear response based approach by Ghosez 
\textit{etal}\cite{GhosezGonze}.
Recently, analysis based on Wannier functions\cite{krak1,krak2} has indicated the important role played by 
the oxygen orbitals in giving rise to large anomalous $Z^\star$ of the TM atoms in TMOs.
However, a quantitative estimation of various contributions such as covalency, local polarizability
and charge transfer to $Z^\star$ is still lacking. 

In this work, we link Wannier function based description with atomic orbitals to quantify
various chemical contributions to $Z^\star$. 
Polarization along $x-$direction is given by the trace of $PxP$, where $x$ is 
the position operator and $P$ an operator of projection into subspace of occupied 
states. A natural microscopic picture of its makeup can be obtained by diagonalization
of the $PxP$ operator. For systems periodic in 1-dimension, eigenfunctions of 
$PxP$ operator are the Wannier functions\cite{BW} (or hermaphrodite orbitals of 3-D
periodic systems\cite{Sgi}). For 3-dimensional periodic systems, $PxP$, $PyP$ and
$PzP$ do not generally commute and can be diagonalized 
simultaneously only approximately, resulting in maximally localized Wannier 
functions\cite{MV}. In analogy with crystal orbital overlap population\cite{hoff},
based on projections of Wannier functions onto atomic orbitals,
we propose here concepts of Wannier Orbital Overlap Population (WOOP) and Wannier Orbital Position 
Population (WOPP), which are used to determine quantitatively the contributions of 
different electronic mechanisms to anomalous dynamical charges. 
In principle, Mulliken\cite{Mull} charge population becomes a subset of WOOP,
if the Wannier functions are replaced by an atomic orbital basis set in which the 
electronic wave functions are expanded. 
On the other hand, WOPP introduces a new scheme to accurately estimate the degree of covalency. 
%

In section \ref{systems}, we describe the systems analyzed in this work.
We review in section \ref{method} the definition of BEC, 
a method to construct localized Wannier Functions (WFs), and introduce WOOP and WOPP.
In Section \ref{R_D}, we present results of our analysis and break-up of the effective
charge into different mechanisms, and finally conclude in section \ref{con}.
%
%
\section{Systems}
\label{systems}

PbTiO$_3$ (PTO) is an end member $(x=1)$ of PbZr$_{1-x}$Ti$_x$O$_3$ (PZT), the most 
commonly used technological ferroelectric material. It occurs in the perovskite
structure with a tetragonal ferroelectric phase up to a relatively high Curie
temperature $T_c= 763$ K, above which it transforms to the cubic paraelectric phase. While the
stereochemical activity of lone pair of $6s$ electrons in Pb is known to be crucial 
for its properties, it shares with other perovskite ferroelectrics the $d^0$ TM-oxygen 
interaction contributing to ferroelectricity. It has been studied extensively
using first-principles calculations\cite{WR}. As indicators of ferroelectric
instability, effective charges of Pb and Ti in its cubic paraelectric phase ($a=3.97$\AA) are 3.9 and 7.1
respectively, which are analyzed here. 

BaTiO$_3$ (BTO), a related material, undergoes a sequence of structural transitions from cubic to
tetragonal, then to orthorhombic and subsequently to rhombohedral phase as temperature is
reduced. In the cubic structure ($a=4.00$\AA), BECs of Ba and Ti are 2.74 and 7.2 respectively.
Polarization of BaTiO$_3$ is relatively small partly because Ba does not have any lone-pair of electrons and
does not contribute much to ferroelectricity. A comparative study of PbTiO$_3$ and BaTiO$_3$
allows us to identify the special role of Pb at $A$-site.

Finally, we analyze effective charges in BaZrO$_3$ (BZO), which remains stable in the paraelectric phase at all 
T, with the cubic perovskite structure ($a=4.19$\AA). Effective charges of Ba and Zr are 2.74 and 5.7 
respectively. Comparison of various contributions to anomalous charges of Zr and Ti allows us
to single out the chemical mechanism which leads to ferroelectricity in BaTiO$_3$, but not
in BaZrO$_3$.

%
\begin{figure}[t]
\centering
\includegraphics[scale=0.30]{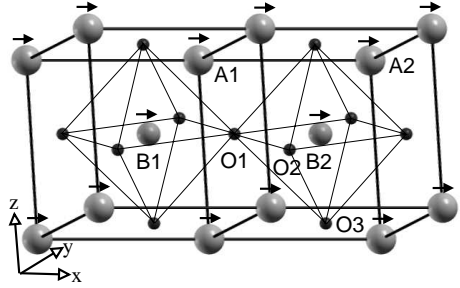}
\caption{Two unit cells of a cubic ABO$_3$ perovskite system.
Labels in this figure are used in discussion.}
\label{str}
\end{figure}
%
\section{Methods}
\label{method}

All calculations reported here have been carried out with 
ABINIT\cite{Abinit} implementation of DFT using Teter extended norm-conserving 
pseudopotentials and a plane wave basis. The $s$ and $p$ semi-core states of Ti, Zr and Ba are 
included in the valence. An energy cutoff of 80 Ry on the plane-wave basis is used 
in representing Kohn-Sham wave functions. Integrals over the Brillouin zone 
have been sampled with a 6$\times$6$\times$6 uniform mesh of k-points. 
Wannier functions have been constructed using post-processing subroutines interfaced
with ABINIT code, developed by the authors\cite{BW}.
%
\subsection{Born effective charge}
\label{BEC_Pol}

Elements of the BECs of an atom (I) are tensors, whose elements $Z^\star_{\alpha\beta,I}$
are the coupling constants between atomic displacements (phonon) and electric field. 
Force $F$ felt by an atom $I$ in the presence of electric field $E$ is: 
\begin{equation}
F_{\alpha, I} = Z^*_{\alpha \beta, I} E_{\beta},
\end{equation}
where as the dipole moment induced by displacement $u$ of an atom is:
\begin{equation}
\label{convdeff}
\Delta \mu_{\alpha} = Z^*_{\alpha \beta, I} u_{I \beta}.
\end{equation}
Corresponding change in the polarization is $\Delta \mu/V$, $V$ being the volume of
unit cell. In this work, we use the relation (\ref{convdeff}) in analyzing the BEC.

A well localized WF $W_n( \mathbf{r} )$ (constructed in the scheme described below) enables evaluation of
expectation values of position operator:
\begin{equation}
<x>_n=\int x W^{\star}_n( \mathbf{r},0)W_n(\mathbf{r},0) d \mathbf{r}. \nonumber
\label{omega}
\end{equation}
The total electronic contribution to the dipole moment per unit cell is obtained 
by summing over all the bands: $\mu_x = e f \sum_n <x>_n$, $e$ being the electronic charge
and $f$ the occupation number of the $n$th band.
The BEC can be determined from the changes in the dipole moment resulting from the off-centering of an atom.
For example, BEC associated with dipole induced along $x$ direction by a small displacement $u_x$ of an ion $I$ 
is given by:
\begin{equation}
Z_{xxI}^\star = Z^{ion}_I + \frac{\Delta \mu^e_x}{u_x},
\label{Born_charge}
\end{equation}
where $Z^{ion}_I$ is the  ionic charge ($Z^{ion}_I =$ number of valence 
electrons of the given atom treated with a pseudopotential).
$\Delta \mu^e_x$ is calculated from the WCs as:
\begin{equation}
\Delta \mu^e_{x}=e \left[{\sum_i^N <x>^{d}_i-\sum_j^N <x>^{nd}_j}\right],
\label{deltap}
\end{equation}
where $<x>^{d}_i$ and $<x>^{nd}_i$ are the expectation values of $x$ obtained using the 
$i$-th WF with and without the atomic displacement respectively. 
WFs thus allow decomposition of the  dynamical charge into contributions from different 
bonding orbitals spanning the occupied subspace. 
\subsection{Wannier functions}
Wannier functions(WF) are obtained as Fourier transform of Bloch functions of occupied bands.
At any wave vector $\mathbf{k}$, any unitary transformation of Bloch functions within the occupied
band subspace is a valid description of electronic structure in the sense that the total 
energy is invariant under such a transformation. As a result, WFs are highly non-unique.
Nevertheless, there are always some special choices of these unitary rotations,
also known as gauge, which yield highly localized sets of WFs\cite{MV,wlo}.

In this work, we use the Wannier type localized orbitals (WLO) described in 
Ref\cite{wlo}:
\begin{equation}
W_I(\mathbf{r}, \mathbf{R})=\sum_{\mathbf{k}} e^{i \mathbf{k}.( \mathbf{r}- \mathbf{R})}\sum_l 
M^\mathbf{k}_{Il} u_{l \mathbf{k}}(\mathbf{r}), \nonumber 
\label{wlo}
\end{equation}
where the wave functions $\left\{ u_{l \mathbf{k}}(\mathbf{r}) \right\}$, obtained from a DFT calculation, are the 
cell periodic part of Bloch functions. Rotation of these functions 
by the unitary matrix $M^\mathbf{k}$ (at each $\mathbf{k}$) is aimed at obtaining a set of wave functions 
that are smooth and periodic in reciprocal space, a criterion necessary to obtain well
localized WFs. 

\begin{figure}[t]
\centering
\includegraphics[scale=0.51]{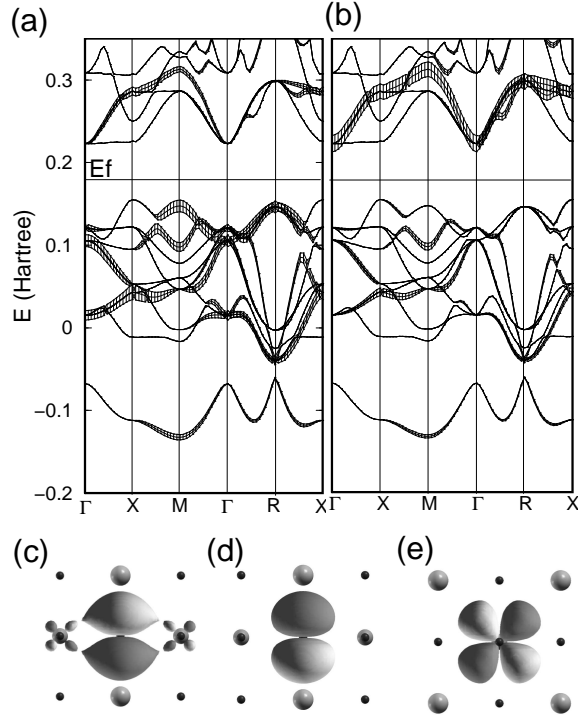}
\caption{Band structure of PbTiO$_3$ with (a) $2p_y$  and (b) $3dxy$ characters centered 
at O1 and Ti respectively.
The $2p_y$ WLO at O1, constructed (c) within the occupied subspace, (d) in an extended
subspace including bands above Ef. (e) The $3d_{xy}$ WLO at Ti constructed in the 
extended subspace.}
\label{fat_bands}
\end{figure}

In ionic systems, WLOs are typically centered on atoms, and their delocalized features
extending to nearest neighbour atoms indicate the degree of covalency. 
In strongly covalent systems, WLOs are typically  bond-centered, while the atom-centered WLOs 
can be obtained by expanding the subspace of electronic states to include the unoccupied, antibonding states.
As demonstrated in Fig-\ref{fat_bands}, the O1-centered [see Fig-\ref{str}] WLO with $2p_y$ orbital character 
and the Ti-centered WLO with $3d_{xy}$ orbital character, are constructed
using an extended subspace above the Fermi level, in cubic PTO. The width or``fatness''\cite{fat_andersen} of the bands in
Fig-\ref{fat_bands}(a) and (b) is linearly proportional to their contribution to the
two WLOs respectively while Fig-\ref{fat_bands}(c) shows the $2p_y$ orbital of O1 constructed 
within the occupied subspace. 
When constructed in the extended subspace, the antibonding contribution 
between the orbitals $2p_y$ of O1 and the $3d_{xy}$ of Ti from the bands above Fermi level
in Fig-\ref{fat_bands}(a) destructively interfere with the corresponding bonding contributions
from the bands below Fermi level to yield a pure $2p_y$ like WLO shown in Fig-\ref{fat_bands}(d). 
Whereas, for the pure Ti $3d_{xy}$, shown in Fig-\ref{fat_bands}(e), the majority contribution
comes from the band above Fermi level, as clearly indicated in Fig-\ref{fat_bands}(b).

In the rest of this paper, we refer to the atom-centered WLOs constructed in the extended subspace
as atomic orbitals (AO) and WLOs constructed within the occupied subspace as WFs. 
  

\subsection{Wannier Orbital Overlap Population (WOOP) and 
Wannier Orbital Position Population (WOPP)}
\label{bectype}
A picture of bonding in a crystal obtained with Wannier functions is qualitative.
To link it quantitatively to different chemical mechanisms of bonding, and 
we now introduce tools based on their projections on atomic orbitals.
For clarity, we denote indices of WFs and atomic orbitals (AO's) in 
upper and lower case letters respectively. 

In the first step, we express the WFs in terms of AOs 
$\left\{ | \phi_j\rangle \right\}$ as: $|W_I \rangle =\sum^N_{j=1} C_{Ij}| \phi_j\rangle$, where the AOs,
$N$ in number, may not necessarily form an orthonormal basis.
$C_{Ij}$'s can be determined by solving $N$ simultaneous linear equations of the form:
\begin{equation}
\sum^N_{j=1} S_{mj} C_{Ij}= \langle \phi_m | W_I \rangle, 
\end{equation}
where $S_{mj}=\langle \phi_m | \phi_j \rangle$.
\begin{figure}[b]
\centering
\includegraphics[scale=0.38]{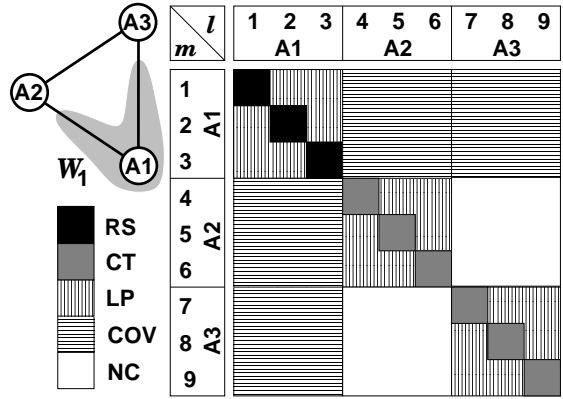}
\caption{A schematic grouping of $l$ and $m$ to decompose $Z^\star$ due to $W_1$(RS: Rigid shift,
CT: Charge transfer, LP: Local polarizability, COV: Covalency, NC: Nonlocal changes).}
\label{boop_scheme}
\end{figure}
Normalization of  $| W_I \rangle$ in terms of AOs is: $\langle W_I | W_I \rangle = 
\sum^N_{lm} C_{Il}^\ast S_{lm} C_{Im}$.
Evidently, a choice of AOs is quite important in obtaining the normalization of WFs
accurately.

The part of the summand corresponding to $l=m$ in the normalization of $| W_I \rangle$ gives the population of electrons at
$| \phi_l \rangle$ due to $| W_I \rangle$. We generalize this to define ``Wannier orbital overlap 
population''(WOOP) associated with the $I$-th WF as:
\begin{equation}
B^I_{lm}= C_{Il}^\ast S_{lm} C_{Im}. 
\label{boop}
\end{equation}
Choosing  $l$ and $m$ corresponding to all the AOs centered on an atom, say A1 in Fig-\ref{boop_scheme},
the population of electrons, which are centered on A1 and contribute to the $W_1$, can be evaluated as:
%
\begin{equation}
Q^1_{A1}= \sum_{lm\in A1} B^1_{lm}. 
\label{epop}
\end{equation}
Subsequently, the number of electrons populated in atom A1 can be estimated as $\sum_I Q^I_{A1}$. 

Center of $| W_I \rangle$ can now be expressed in terms of the AOs:
$\langle x\rangle_I= \sum_{lm} C_{Il}^\ast X_{lm}C_{Im}$, where  $X_{lm}= \langle \phi_l |x| \phi_m \rangle $.
We refer to the summand as ``Wannier orbital position population''(WOPP) and denote it as:
\begin{equation}
D^I_{lm} = C^{\ast}_{Il} X_{lm} C_{Im}.
\label{bold}
\end{equation}
This leads to the net contribution to $Z^\star$ from $| W_I \rangle$ as: 
\begin{equation}
Z^{\star I}_x = \frac{e}{u_x} \sum^N_{l,m=1} \left( D^{dI}_{lm} - D^{ndI}_{lm} \right).
\label{zi}
\end{equation}
Through appropriate choices of $l$ and $m$ in eqn-(\ref{zi}), the $Z^{\star I}_x$ can be easily 
decomposed into  contributions based on different chemical mechanisms responsible.
These are classified into five broad categories: rigid shift (RS), 
charge transfer (CT)\cite{GhosezGonze}, local polarizability (LP), covalency (COV) and 
other nonlocal changes (NC). For the WF centered on A1 (see schematic in Fig-\ref{boop_scheme}), 
these are defined by the following conditions on $l$ and $m$:
\begin{eqnarray}
\label{breakup}
&1.&\mbox{RS: } l=m, \left\{ lm \right\} \in \mbox{Ai}, \Delta Q^1_{Ai} =     0, i=1,2,3; \nonumber\\
&2.&\mbox{CT: } l=m, \left\{ lm \right\} \in \mbox{Ai}, \Delta Q^1_{Ai} \not= 0, i=1,2,3; \nonumber\\
&3.&\mbox{LP: } l\not=m, \left\{ lm \right\} \in \mbox{A}; \nonumber\\
&4.&\mbox{COV: } l\not=m, l\in  \mbox{A},m \not\in  \mbox{A} 
   \mbox{ or }m\in  \mbox{A},l \not\in  \mbox{A}; \nonumber\\
&5.&\mbox{NC: } l\not=m, \left\{ lm \right\} \not\in \mbox{A}.
\end{eqnarray}

RS and CT are contributed from the diagonal blocks in Fig. \ref{boop_scheme} 
(identical AOs) and can be differentiated based on the change in the population ($\Delta Q$), obtained through WOOP.
Whereas,  $Z^\star$ from the LP arises solely from the coupling between the AOs with 
different parities centered on the same atom, for which the inter orbital transition dipole moment is nonzero. 
Due to orthonormality of AOs centered on the same atom, contribution of LP to $Z^\star$ is not 
effected by the rigid translation of AOs. 
  
Contributions to $Z^\star$ from COV and NC are both due to transition dipole moments
between orbitals centered on different atoms.
In Fig-\ref{boop_scheme} where $|w1\rangle$ has highest projection with AOs of A1, COV gives 
covalent contribution to $Z^\star$ due to transition dipole moments between the AOs of A1 and
it's nearest neighbours A2 and A3. COV contributions can be both bonding
as well as antibonding type in nature. 
Finite COV as well as CT contributions to BEC, involving
same set of AOs of two neighbouring atoms, can be indicative of a possible conjugation scenario, 
as shown in section-\ref{R_D}.
NC gives the contribution to $Z^\star$ from covalency associated with transition dipole moments 
between the AOs of A2 and A3[Fig-\ref{boop_scheme}]. 
In general, NC contribution to $Z^\star$ is much smaller than the other four categories
as none of the AOs involved in NC belong to A1. 

\begin{table}[b]
\caption{Various chemical contributions to Z$^\star_{Ti}$ and Z$^\star_{Pb}$ in PbTiO$_3$.(CP:Charge transfer; 
RS:Rigid shift; LP:Local polarizability; COV:Covalency; NC:Non-local changes)}
\begin{center}
\begin{tabular*}{5.0in}{c@{\extracolsep{\fill}}c@{\extracolsep{\fill}}c@{\extracolsep{\fill}}
c@{\extracolsep{\fill}}c@{\extracolsep{\fill}}c@{\extracolsep{\fill}}}
\hline 
\hline 
         WF   & CT   & LP    & COV  & NC   & Tot  \\
\hline 
\multicolumn{6} {c}{Decomposition of Z$^\star_{Ti}$ from orbitals of O1}\\
\hline
$2s+2p_x   $ & 0.2l & -0.10 & 0.41 & 0.01 & 0.53  \\ 
$2s-2p_x   $ & 0.33 & -0.06 & 0.30 & 0.01 & 0.58  \\   
$2p_y,2p_z $ & 1.10 &  0.00 & 0.27 & 0.06 & 1.43  \\   
\hline 
\multicolumn{6} {c}{Decomposition of Z$^\star_{Pb}$ from orbitals of O2}\\
\hline
$2p_x$       & 0.12 & -0.08 & 0.28 & 0.02 & 0.34       \\
$2p_y\pm 2s$ & 0.02 & -0.04 & 0.03 & -0.00 & 0.01 \\
$2p_z$       & 0.38 & -0.03 & 0.07 &  0.01 & 0.43    \\ 
\hline
\multicolumn{6} {c}{Decomposition of Z$^\star_{Pb}$ from orbitals of Pb}\\
\hline  
         WF   & RS   & LP    & COV  & NC   & Tot  \\
\hline
$6s$  & -1.97  & 0.58 & -0.02 & -0.02 & -1.43  \\ 
\hline 
\hline
\end{tabular*}
\end{center}
\label{PTO}
\end{table}
\begin{figure}[b]
\centering
\includegraphics[scale=0.5]{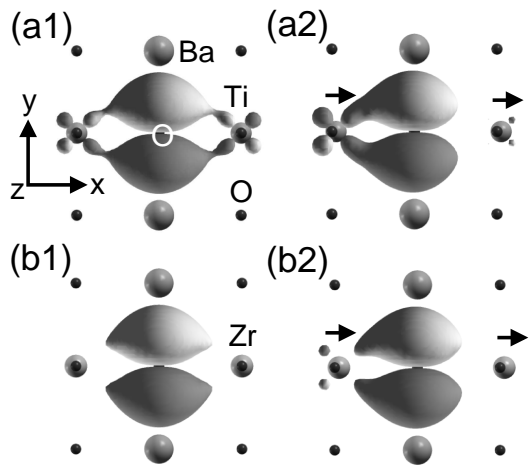}
\caption{O1-centered WFs with symmetry of its $2p_y$ orbital in cubic (a) BaTiO$_3$ and (b) BaZrO$_3$
(1)with and (2) without Ti/Zr displacement. The isosurfaces shown here, are for $|W(\mathbf{r})|=0.0045$, 
which is typically about 10\% of its maximum.}
\label{Tidisp}
\end{figure}
\begin{figure}[b]
\centering
\includegraphics[scale=0.5]{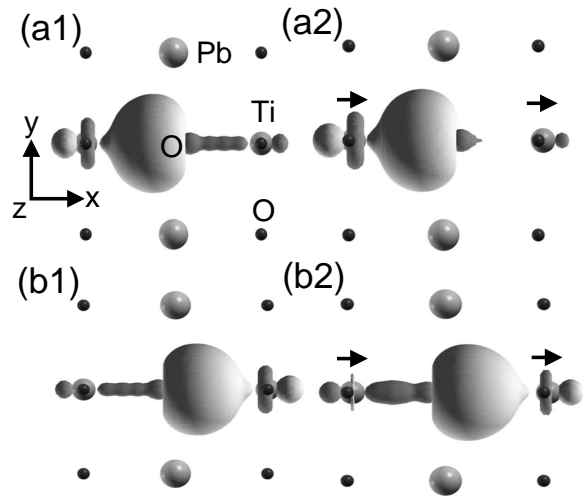}
\caption{O1-centered WFs with symmetries of (a) $2s+2p_x$ and (b) $2s-2p_x$ orbitals 
(1) with and (2) without Ti displacement in cubic PbTiO$_3$. The isosurfaces are 
for $|W(\mathbf{r})|=0.0045$, which is about 10\% of its maximum.}
\label{Tidisp2px}
\end{figure}
\begin{table}[b]
\caption{Various chemical contributions to Z$^\star_{Ti}$ and Z$^\star_{Ba}$ in BaTiO$_3$.}
\begin{center}
\begin{tabular*}{5.0in}{c@{\extracolsep{\fill}}c@{\extracolsep{\fill}}c@{\extracolsep{\fill}}
c@{\extracolsep{\fill}}c@{\extracolsep{\fill}}c@{\extracolsep{\fill}}}
\hline 
\hline 
      WF   & CT     & LP       & COV    & NC      & Tot  \\
\hline 
\multicolumn{6} {c}{Decomposition of Z$^\star_{Ti}$ from orbitals of O1}\\
\hline
$2s+2p_x   $ & 0.19 & -0.08 & 0.42 & -0.01 & 0.52 \\   
$2s-2p_x   $ & 0.34 & -0.05 & 0.29 & -0.00 & 0.58 \\   
$2p_y,2p_z $ & 1.18 & -0.01 & 0.26 &  0.06 & 1.49 \\   
\hline
\multicolumn{6} {c}{Decomposition of Z$^\star_{Ba}$ from orbitals of O2}\\
\hline
$2p_x $      & -0.01 & -0.05 & 0.31 &  0.01 & 0.26 \\   
$2s\pm2p_y $ &  0.04 & -0.04 & 0.06 & -0.00 & 0.06 \\   
$2p_z $      &  0.11 &  0.00 & 0.11 & -0.00 & 0.22 \\   
\hline
\multicolumn{6} {c}{Decomposition of Z$^\star_{Ba}$ from orbitals of Ba}\\
\hline
      WF   & RS     & LP       & COV    & NC      & Tot  \\
\hline
$5p_x $      & -1.94 & 0.08 & -0.02 & 0.00 & -1.88  \\
$5p_y,5p_z $ & -1.94 & 0.09 & -0.00 & 0.00 & -1.85  \\   
\hline 
\hline
\end{tabular*}
\end{center}
\label{BTO}
\end{table}
\begin{table}[b]
\caption{Various chemical contributions to Z$^\star_{Zr}$ and Z$^\star_{Ba}$ in BaZrO$_3$.}
\begin{center}
\begin{tabular*}{5.0in}{c@{\extracolsep{\fill}}c@{\extracolsep{\fill}}c@{\extracolsep{\fill}}
c@{\extracolsep{\fill}}c@{\extracolsep{\fill}}c@{\extracolsep{\fill}}}
\hline \hline 
      WF   & CT    & LP    & COV  & NC   & Tot  \\
\hline
\multicolumn{6} {c}{Decomposition of Z$^\star_{Zr}$ from orbitals of O1}\\
\hline
$2s+2p_x   $ &-0.16 & -0.10 & 0.63 & 0.00 & 0.37  \\   
$2s-2p_x   $ &-0.01 & -0.10 & 0.50 & 0.01 & 0.40  \\  
$2p_y,2p_z $ & 0.89 & -0.04 & 0.22 & 0.03 & 1.10 \\   
\hline
\multicolumn{6} {c}{Decomposition of Z$^\star_{Ba}$ from orbitals of O2}\\
\hline
$2p_x $      & 0.03 & -0.03 &  0.23 & 0.02 &  0.25     \\   
$2s\pm2p_y $ & 0.04 & -0.03 & 0.04 & -0.00 & 0.05  \\ 
$2p_z $      & 0.13 & 0.00 & 0.07 & -0.01 & 0.19  \\   
\hline
\multicolumn{6} {c}{Decomposition of Z$^\star_{Ba}$ from orbitals of Ba}\\
\hline
      WF   & RS     & LP       & COV    & NC      & Tot  \\
\hline
$5p_x $      & -1.93 & 0.07 & -0.03 & -0.00 & -1.89  \\
$5p_y,5p_z $ & -1.94 & 0.07 & -0.00 & -0.00 & -1.87   \\   
\hline
\hline
\end{tabular*}
\end{center}
\label{BZO}
\end{table}
\begin{table}[t]
\caption{Comparison of various contributions to anomalous effective charges.}
\begin{center}
\begin{tabular*}{5.0in}{c@{\extracolsep{\fill}}c@{\extracolsep{\fill}}c@{\extracolsep{\fill}}
c@{\extracolsep{\fill}}c@{\extracolsep{\fill}}}
\hline 
\hline 
              & Charge & Covalency & \multicolumn{2} {c}{Local} \\
              &  transfer  &    & \multicolumn{2} {c}{polarizability} \\
\hline 
 & \multicolumn{4} {c} {(from orbitals of O1)} \\
Ti (BTO)      & 2.89  &  1.23 &  \multicolumn{2} {c}{-0.15} \\
Zr (BZO)      & 1.61  &  1.57 &  \multicolumn{2} {c}{-0.28} \\
\hline
   & \multicolumn{2}{c}{(O2+O3+A)}  & (O2+O3)& (A) \\
Ba (BTO)      & 0.36  &  1.06  &  -0.26 & 0.26\\
Pb (PTO)      & 1.08  &  0.80  &  -0.38 & 0.58 \\
\hline 
\hline 
\end{tabular*}
\end{center}
\label{sum}
\end{table}
\begin{figure}[t]
\centering
\includegraphics[scale=0.4]{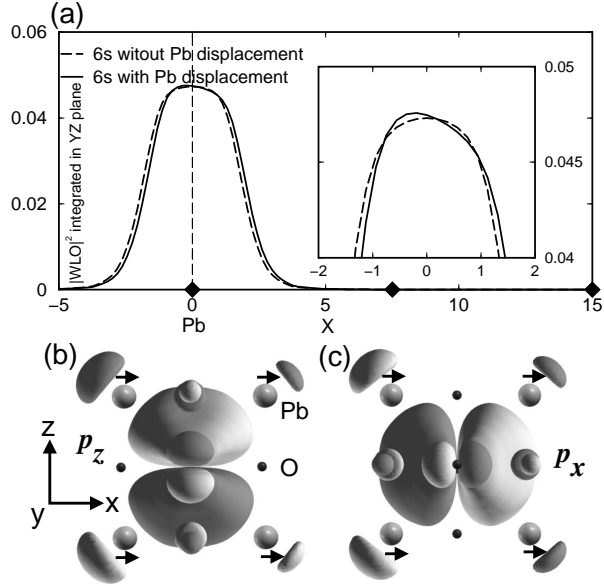}
\caption{WFs with symmetry of (a) $6s$ orbital of Pb, (b) $2p_z$ and (c) $2p_x$ orbitals of O1
in cubic PbTiO$_3$. Isosurfaces shown in (b) and (c) are for $|W(\mathbf{r})|=0.001$.}
\label{Pbdisp}
\end{figure}
\begin{figure}[t]
\centering
\includegraphics[scale=1.75]{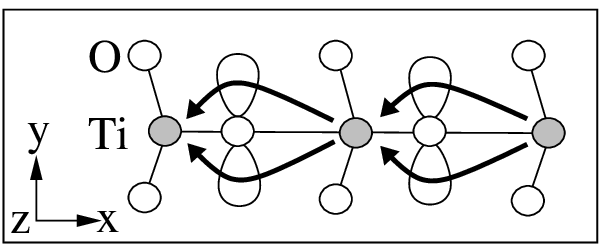}
\caption{A schematic picture of interatomic charge transfer across one unit cell.}
\label{CT_schematic}
\end{figure}


\section{Results and discussion}
\label{R_D}


We first present WOPP based analysis of $Z^\star_{Ti}$ in PTO and compare it with
that of $Z^\star_{Ti}$ in BTO. 
Next, we analyze $Z^\star_{Pb}$ of PTO and bring out the special
role played by Pb through comparison with Ba. Finally, we present a comparative analysis 
of BTO and BZO to identify factors crucial to ferroelectricity in BTO and to the 
lack of it in BZO. In all these cases, cations are off-centered
(with respect to the cubic structure) along $x-$axis by 1 \% of the lattice constant.

In addition to a nominal charge of 4 a.u., the anomalous part of $Z^\star_{Ti}$ in PbTiO$_3$ is
about 3.1 a.u. To this anomalous part, a positive contribution of 4.2 a.u. is
from the orbitals of O1, which belongs to the Ti-O chain along the $x$-axis, and a negative contribution of 
-1.1 a.u.  from the orbitals of O2 and O3 in the Ti-O plane perpendicular to the $x$-axis. Almost 70\% 
of the total positive contribution (4.2 a.u.) is made by the $2p_y$ and $2p_z$ orbitals of O1 
(both are equal by symmetry), which are perpendicular to the Ti-O chain along $x$-axis. 
The O1-centered WFs with $2p_y$ orbital character are shown in Fig-\ref{Tidisp}. 
The majority of their contributions (see Table-\ref{PTO})
to $Z^\star_{Ti}$ is due to  inter-atomic charge transfer across the unit cell from the $3d_{xy}$ and 
$3d_{xz}$ orbitals of Ti1 to those of Ti2. WOOP analysis for these WFs clearly shows a decrease and 
increase in electron population at Ti1 and Ti2 
respectively, with negligible change in the population of orbitals at O1.

Bonding in PTO is certainly not purely ionic: a small but non-zero covalent contribution of about 0.3 a.u.
to $Z^\star$ from mixing between the O1 and Ti orbitals indicates a weak $\pi$ like bonding 
along the  Ti1-O1-Ti2 chain which, is strengthened in the ferroelectric phase. 
The rest 30\% of the total positive contribution to the anomalous part of $Z^\star_{Ti}$ is 
due to the $2s\pm2px$ hybridized orbitals of O1. Unlike the 2$p_y$ and 2$p_z$ orbitals 
which are perpendicular to the -O-Ti-O- chain direction, covalent contributions from these orbitals
arises from $\sigma$ type bonding between Ti and O and are comparable to that from charge transfer.
These WFs (shown in Fig-\ref{Tidisp2px}) are centered between O and Ti atoms: 
the $2s+2p_x$ orbital is peaked in the bond path between O1 and Ti1 where as
$2s-2p_x$ peaked is peaked in the bond path between O1 and Ti2. The former 
contracts and the latter gets stretched in the ferroelectric phase. 
Consequently, in the ferroelectric 
phase the covalency between O1 and Ti1 gets stronger at the cost of the same between O1 and Ti2.
This is similar to $\pi$ conjugation found in long chain conducting polymers.
As shown in Table-\ref{PTO}, this results into a larger covalent type contribution to 
$Z^\star_{Ti}$ from $2s+2p_x$ than that from $2s-2p_x$ WF. 
These orbitals too facilitate the transfer of small electronic charge from Ti2 
to Ti1, resulting in charge transfer contributions to $Z^\star_{Ti}$ that are about the same
as their contribution with covalency.
Small contributions from local polarizability of oxygen to $Z^\star_{Ti}$ from these orbitals 
correlate with the intra-atomic change in hybridization among the $s,p$ and $d$ orbitals of oxygen without 
altering the total charge population or WOOP at O1. Similarly, local polarizability of O2
(and changes in its hybridization) yield a small 
negative contributions to $Z^\star_{Ti}$.

Interestingly, charge transfer between TM atoms at $B$-site through O seems to be indirectly
influenced by the atom A (or its size). Comparison of contributions to $Z^\star_{Ti}$ in PTO and BTO
(shown in Tables-\ref{PTO} and \ref{BTO}) indicates that charge transfer from Ti2 to Ti1 is higher 
in BaTiO$_3$ than in PbTiO$_3$. In contrast, the
contributions from the $2s\pm2p_x$ hybridized orbitals ($\sigma$ bonding) of O1 in BaTiO$_3$  
are about the same as in PbTiO$_3$. As we will see below, the former difference is due to the fact
that Pb also competes in contributing to the charge transfer through the same O1 orbitals.

Anomalous part of $Z^\star_{Pb}$ in PbTiO$_3$ is about 1.9 a.u., over a nominal charge of 2.0 a.u.
Almost 25\% of this anomalous part is a result of local polarizability at Pb due to
emergence of moderate Pb $6p$ orbital character in occupied states of the ferroelectric phase,
which is accompanied by a small decrease in occupancy of the $6s$ orbital. 
This is evident from the nature of deformation of the Pb-centered 
WF of $s$ symmetry [see Fig-\ref{Pbdisp}(a)] in the ferroelectric phase.
The rest of the anomalous part is contributed by the two oxygens O2 and O3 in the (100) 
Ti-O plane. $2p_x$ and $2p_z$ symmetry WFs centered on both of these oxygen atoms
give positive contributions to $Z^\star_{Pb}$. 
Interaction between Pb and O2 orbitals is evident in the long-range features of 
the $2p_x$ and $2p_z$ like WF centered on O2 (shown in Fig-\ref{Pbdisp}(c) and (b))
in the ferroelectric phase. 
The contribution from $2p_x$ is primarily covalent in nature involving 
a combination of $6p$ orbitals of Pb, while the same from $2p_z$ 
is mainly due to charge transfer from similar orbitals of Pb2 to that of Pb1 (see Table-\ref{PTO}).   
 
In both BaTiO$_3$ and BaZrO$_3$, the anomalous part of $Z^\star_{Ba}$ is about 0.75 a.u.
The $5d$ orbitals of Ba, which are unoccupied in the paraelectric phase, mix with 
occupied states in the ferroelectric phase giving a positive contribution to $Z^\star_{Ba}$. 
These orbitals are relatively more localized than the 6s or 6p orbitals of Pb and hence 
less favourable for inter-atomic orbital interactions. 
Consequently, reduction in both charge transfer as well as covalent 
contributions involving the $2p$ orbitals of O2 and O3 and the $5d$ orbitals of Ba is expected.
Similarly, the local polarizability contribution from Ba due to hybridization between its $5p$ and  $5d$ 
orbitals in the ferroelectric phase is also weaker. 
These are evident from comparison between contributions to $Z^\star_{Ba}$ (Tables-\ref{BTO} and \ref{BZO})
and those to $Z^\star_{Pb}$ (Table-\ref{PTO}). 
Our analysis of $Z^\star_{Zr}$ (Table-\ref{BZO}) reveals two interesting
trends: (a) contribution from covalency between Zr and O 
($\sigma$-type bonding) to anomalous part of $Z^\star_{Zr}$ is greater than
that from covalency between Ti and O to $Z^\star_{Ti}$ and (b) charge transfer
is much more significant in giving anomalous $Z^\star_{Ti}$ than in
the case of  $Z^\star_{Zr}$.

We sum up contributions from all the WFs and summarize the decomposition of 
anomalous $Z^\star$'s of A and B cations into various chemical mechanisms (see Table-\ref{sum}).
Comparison between BZO and BTO clearly reveals the role of long-range charge transfer between
TM cations via $p$ orbitals of oxygen (see Fig-\ref{CT_schematic}). This is analogous to 
super-exchange in magnetic oxide materials. As expected from the fact that 
the radius of Zr$^{4+}$ is larger than that of Ti$^{4+}$, contribution to $Z^\star$ from covalency
is greater for Zr than for Ti.
In case of $Z^\star_{Pb}$, the net charge transfer through the orbitals of O(2,3) is the
prime contributor, followed by the net contribution from covalency between the orbitals of O(2,3)
and Pb. The local polarizability of the 6$s$ orbital of Pb also contributes 
sizeably to $Z^\star_{Pb}$, but is partly compensated by the local polarizability of oxygen.
The cause for difference between $Z^\star_{Pb}$ and $Z^\star_{Ba}$ is most evident in
charge transfer contribution, and to a lesser extent in contributions from their local
polarizability and covalency with oxygen.
\section{Conclusion}
\label{con}
We presented a simple scheme to quantify different chemical bonding mechanisms through projections
of Wannier functions on to atomic orbitals. This is particularly useful in separating
the total BEC into various chemical contributions such as local polarizability, covalency and charge transfer.
We find that anomalously large effective charge of Ti in BaTiO$_3$ and PbTiO$_3$ arises primarily
from a dynamical transfer of \textit{tiny} electronic charge through one unit cell from $d$ orbitals of one
Ti atom to its nearest neighbour Ti atom
 in the direction opposite to Ti displacement and secondarily from covalency (the
latter is less than half of the former). 
In contrast, the anomalous part of effective charge of Pb has comparable 
contributions from charge transfer and covalency, of which the former is lacking
in the case of Ba.

We have identified the {\it dynamical charge transfer} between cations (see Fig.-\ref{CT_schematic})
(TM as well as the one at $A$-site) as the key factor for their potential to drive
ferroelectricity. As this charge transfer occurs along -TM-O- chains through
the $p$ orbitals of oxygen that are perpendicular to the chains, it is clear that
perovskite is a structure very special for ferroelectricity for two reasons:
(a) it has -TM-O- chains running in all the three directions and (b) its TMO$_6$ 
octahedra are all corner-shared with two-fold coordination of oxygen atoms giving 
two of the $p$ orbitals of oxygen for facilitating this charge transfer. 
More generally, the tools of WOOP and WOPP developed here, will be useful in characterization of 
chemical bonding in variety of difference systems and phenomena.

\section{Acknowledgments}
JB thanks CSIR, India for a research fellowship and UVW acknowledges a DuPont Young Faculty grant that 
supported some of this work. We are grateful for use of the central computing facility at JNCASR,
funded by the Department of Science and Technology, Government of India. 


\end{document}